\documentclass{svjour3}                     
\smartqed  
\usepackage{graphicx}
\usepackage{amsmath,amsfonts,amsbsy,amssymb,fullpage,makeidx,array,graphicx,epsfig,enumerate}

\usepackage{picinpar}

 \newcommand{\df}{\ {\overset {\rm def} =}\ }
 \newcommand{\dr}[2]{\frac {{\rm d} {#1}} {{\rm d} {#2}}}

\begin{document}

\title{The charged dust solution of Ruban
-- \\
matching to Reissner--Nordstr\"{o}m and shell crossings}

\titlerunning{Charged dust solution of Ruban}

\author{Andrzej Krasi\'nski
  \and
Gabriel Giono
}

\institute{Andrzej Krasi\'nski \\
N. Copernicus Astronomical Centre, Polish Academy of Sciences,
Bartycka 18, 00 716 Warszawa, Poland, \\
           email: akr@camk.edu.pl \\
 \and
 Gabriel Giono \\
D\'epartement de Physique, Universit\'e Claude Bernard Lyon 1, 14 rue Enrico Fermi
69622 Villeurbanne, France, \\
           email: gabriel.giono@etu.univ-lyon1.fr \\
}

\maketitle

\begin{abstract}
The maximally extended Reissner--Nordstr\"{o}m (RN) manifold with $e^2 < m^2$
begs for attaching a material source to it that would preserve the infinite
chain of asymptotically flat regions and evolve through the wormhole between the
RN singularities. So far, the attempts were discouraging. Here we try one more
possible source -- a solution found by Ruban in 1972 that is a charged
generalisation of an inhomogeneous Kantowski--Sachs-type dust solution. It can
be matched to the RN solution, and the matching surface must stay all the time
between the two RN event horizons. However, shell crossings do not allow even
half a cycle of oscillation between the maximal and the minimal size.
\end{abstract}

\maketitle

\section{Motivation}

Avoiding the Big Bang singularity in cosmological models had been a recurring
idea in the literature. The hopes for constructing a model without singularity
were largely dashed by the singularity theorems of Hawking and Penrose (see Ref.
\cite{HaEl1973} for a review). They were temporarily revived by the finding of
Vickers \cite{Vick1973} that in a charged dust model generalising that of
Lema\^{\i}tre \cite{Lema1933} and Tolman \cite{Tolm1934} (LT) the Big Bang can
be prevented by the charge distribution,\footnote{This charged generalisation of
the LT model was first found as a solution of the Einstein -- Maxwell equations
by Markov and Frolov in 1970 \cite{MaFr1970}.} provided that the absolute value
of the charge density is, in geometrical units, \textit{smaller} than the mass
density. This finding created another expectation, that a finite ball of charged
dust matched to the Reissner \cite{Reis1916} -- Nordstr\"{o}m \cite{Nord1918}
(RN) solution would be able to collapse and bounce through the wormhole of the
maximally extended RN manifold, thereby giving physical meaning to the infinite
chain of black holes and asymptotically flat regions.\footnote{The extension of
the RN solution composed of two coordinate patches was first calculated by
Graves and Brill in 1960 \cite{GrBr1960}. The infinite mosaic of conformal
diagrams shown in Fig. \ref{reinormax} first appeared in the paper by Carter
\cite{Cart1966} in 1966, and was described in detail in a review article by
Carter \cite{Cart1973}. See Ref. \cite{PlKr2006} for another pedagogical
introduction.} However, this renewed hope was dashed again in 1991 by Ori
\cite{Ori1991} who proved that, under exactly the same conditions that prevent
the Big Bang, shell crossings will inevitably appear and destroy the dust ball
before it enters the wormhole. Then, Krasi\'nski and Bolejko
\cite{KrBo2006,KrBo2007} found a gap in Ori's assumptions and tried to improve
upon his result. Namely, Ori assumed that the ratio of the absolute value of the
charge density to the mass density (call this ratio $\alpha$) is smaller than 1
everywhere, including the centre of symmetry. Krasi\'nski and Bolejko considered
the case when $\alpha \to 1$ at the centre, while being $< 1$ everywhere else.
The situation turned out to be better, but not definitively. With initial
conditions carefully tuned, the charged dust ball could go through the wormhole
just once, being destroyed by shell crossings soon after reaching the maximal
size in the next asymptotically flat region. Moreover, a direction-dependent
singularity necessarily appeared at the centre of the ball at the instant of
minimal size. This is not a satisfactory situation, but the best result achieved
so far.

In an attempt to find a better model, we now tried to match the Ruban
\cite{Ruba1972} charged dust solution to the RN metric and see what results. The
charged Ruban solution is a generalisation to nonzero charge of the nonstatic
dust solution investigated earlier by Ruban \cite{Ruba1968,Ruba1969}, which in
turn is an inhomogeneous generalisation of the Kantowski -- Sachs model
\cite{KaSa1966}.\footnote{The neutral dust solution investigated by Ruban was
first found as a solution of the Einstein equations by Datt in 1938
\cite{Datt1938}, but instantly dismissed as being of ``little physical
significance''.} The matching of these two solutions is very easily achieved,
with an interesting result. The Ruban solution represents a pulsating charged
dust ball whose outer surface must remain forever between the two event horizons
of the RN solution. It touches the inner event horizon at its minimal size and
the outer horizon at its maximal size. However, a simple investigation shows
that shell crossings are inevitable inside the ball and do not allow even a half
cycle of an oscillation, appearing between the maximum and minimum size both in
the expansion phase and in the collapse phase.

Thus, the field is still open for finding a physical nonstatic source for the RN
solution that would proceed through the wormhole. The next thing to do is to
find a charged perfect fluid solution of the Einstein -- Maxwell equations in
which pressure gradients would prevent the formation of shell crossings. This,
however, promises to be an extremely difficult task.

In this paper, we first present the RN solution in coordinates adapted to the
matching, then the charged Ruban solution as found by the author, then we prove
that the matching can be done, and finally we prove that shell crossings in the
Ruban solution are inevitable. We also investigate the limit of zero charge of
the Ruban -- RN configuration. Then the dust source is matched to the
Schwarzschild solution across a hypersurface that remains inside the
Schwarzschild horizon, only touching it from inside at the moment of maximal
expansion. Shell crossings can then be avoided by an appropriate choice of the
arbitrary functions in the Ruban model. However, this model has a finite time of
existence, being born in a Big Bang that matches to the past Kruskal
\cite{Krus1960} -- Szekeres \cite{Szek1960} singularity and crushing into a Big
Crunch that matches to the future KS singularity.

We use the signature $(+ - - -)$. We do not assume ``units in which $G = c
=1$''; whenever these constants are absent, this means they were absorbed into
other symbols. For example, our time coordinate will be $t = c \times$ [the
physical time], our $M$ will denote $(G / c^2) \times$ [the mass], etc. The
labelling of the coordinates is $(x^0, x^1, x^2, x^3) = (t, r, \vartheta,
\varphi)$.

\section{The Reissner--Nordstr\"{o}m solution with $e^2 < m^2$ and its maximal
extension}

\setcounter{equation}{0}

\begin{figure}
 \begin{center}
 \hspace*{20mm}
 \includegraphics[scale=0.6]{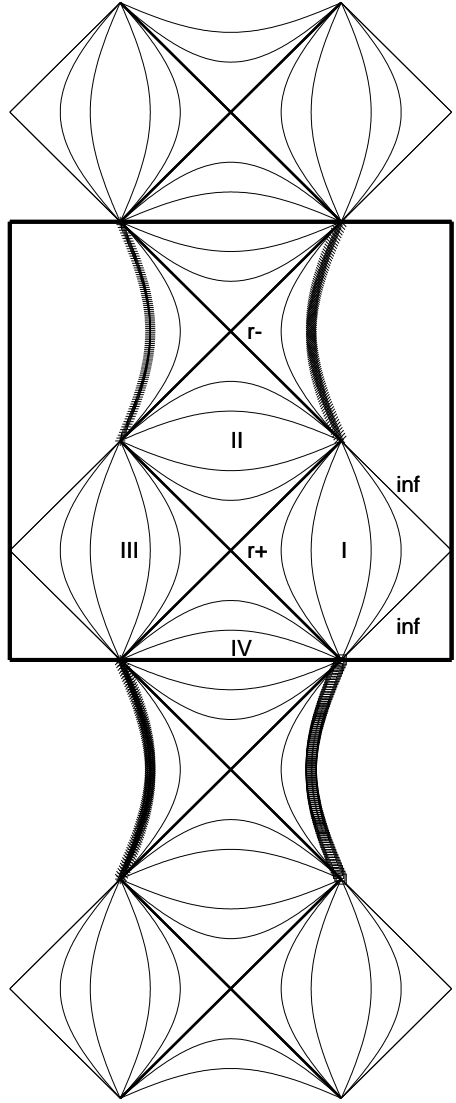}
 \caption{
 \label{reinormax}
 \footnotesize
The conformal diagram of the maximally extended Reissner--Nordstr\"{o}m
spacetime with $\Lambda = 0$ and $e^2 < m^2$. It can either be interpreted as an
infinite chain of copies of the segment within the rectangle, or one can
identify the upper side of the rectangle with the lower side, obtaining a
manifold with closed timelike and null lines. More explanation in the text.}
 \end{center}
 \end{figure}

The Reissner \cite{Reis1916} -- Nordstr\"{o}m \cite{Nord1918} solution in its
standard form is
\begin{equation}
{\rm d}s^2 = F {\rm d}t^2 - (1 / F) {\rm d}r^2 - r^2 \left({\rm d}\vartheta^2 +
\sin^2 \vartheta {\rm d} \varphi^2\right) \label{2.1}
\end{equation}
where
\begin{equation}
F(r) \df 1 - \frac {2m} {r} + \frac{e^2}{r^2} + \frac{1}{3}\Lambda r^2
\label{2.2}
\end{equation}
(we include $\Lambda$ for a while). For reference, Fig. \ref{reinormax} (adapted
from Ref. \cite{PlKr2006}) shows the Carter -- Penrose diagram of the maximal
analytic extension of the underlying spacetime for the case $\Lambda = 0$ and
$e^2 < m^2$, to which we will limit our attention in the main part of this
paper. The ragged lines in the diagram are the curvature singularities at $r =
0$, the lines marked ``inf'' are the past and future null infinities, $r_+$ are
the outer event horizons at
\begin{equation}\label{2.3}
r = r_+ = m + \sqrt{m^2 - e^2},
\end{equation}
and $r_-$ are the inner event horizons at
\begin{equation}\label{2.4}
r =  r_- = m - \sqrt{m^2 - e^2}.
\end{equation}
These two values of $r$ are zeros of the function $F$ when $\Lambda = 0$.
Regions I and III are asymptotically flat, regions II and IV are contained
between the two horizons. The thin curved lines are those on which $r$ is
constant; they are timelike in the asymptotically flat regions where $r > r_+$
and inside the inner event horizon where $r < r_-$; between the horizons where
$r_- < r < r_+$ they are spacelike.

In the following we will be interested in the region between the horizons, where
$F < 0$, so $r$ in (\ref{2.1}) becomes the time coordinate and $t$ becomes a
spacelike coordinate. It is easy to verify that the lines on which $(t,
\vartheta, \varphi)$ are all constant are geodesics.

For later convenience, we rename the coordinates in (\ref{2.1}) as follows
\begin{equation}\label{2.5}
(t, r) = (\rho, R),
\end{equation}
and then transform the $R$-coordinate in the region $F < 0$ to $\tau(R)$ defined
by
\begin{equation}\label{2.6}
\left(\dr R {\tau}\right)^2 = - F = - 1 + \frac {2m} {R} - \frac {e^2} {R^2} -
\frac{1}{3}\Lambda R^2.
\end{equation}
After this, the metric in this region becomes
\begin{equation}\label{2.7}
{\rm d} s^2 = {\rm d} \tau^2 - [- F(R(\tau))] {\rm d} \rho^2 - R^2(\tau)
\left({\rm d}\vartheta^2 + \sin^2 \vartheta {\rm d} \varphi^2\right),
\end{equation}
with $R(\tau)$ being defined by (\ref{2.6}).

When $\Lambda = 0$, eq. (\ref{2.6}) can be integrated to give the following
explicit function $\tau(R)$:
\begin{equation} \label{2.8}
\tau - \tau_0 = \mu \int \frac {{\rm d}R} {\sqrt{-1 + {\displaystyle{\frac {2m}
R}} - {\displaystyle{\frac {e^2} {R^2}}}}} = - \mu \sqrt{- R^2 + 2mR - e^2} +
\mu m \arcsin \left(\frac {R - m} {\sqrt{m^2 - e^2}}\right),
\end{equation}
where $\mu = \pm 1$ and $\tau_0$ is an arbitrary constant.

For some purposes it is more convenient to introduce the parameter $\eta$ by $(R
- m) / \sqrt{m^2 - e^2} = \sin (\eta - \pi/2) \equiv - \cos \eta$, and then
\begin{equation}\label{2.9}
R = m - \sqrt{m^2 - e^2} \cos \eta, \qquad \tau - \widetilde{\tau}_0 = \mu
\left(m \eta - \sqrt{m^2 - e^2} \sin \eta\right),
\end{equation}
where $\widetilde{\tau}_0 = \tau_0 - \mu m \pi/2$. This shows that as $\tau$
increases, $R$ is oscillating between the $r_-$ and $r_+$ given by (\ref{2.3})
and (\ref{2.4}).

\section {The charged Ruban solution \cite{Ruba1972}}

\setcounter{equation}{0}

To derive this solution, we assume comoving coordinates, spherical symmetry and
a charged dust source. For the metric we assume a less general form than
spherical symmetry would allow, namely
\begin{equation}\label{3.1}
{\rm d} s^2 = {\rm e}^{C(t,r)} {\rm d} t^2 - {\rm e}^{A(t, r)} {\rm d} r^2 -
R^2(t)\left[{\rm d} \vartheta^2 + \sin^2(\vartheta) {\rm d} \varphi^2\right],
\end{equation}
where $C(t,r)$, $A(t,r)$ and $R(t)$ are functions to be found from the Einstein
-- Maxwell equations. The limitation of generality is in $R(t)$; in the most
general case it would depend on $r$ as well and the field equations would lead
to a charged generalisation of the LT model.

Details of the calculation are given in Ref. \cite{PlKr2006}. The field
equations show that $R,_t C,_r = 0$. With $R,_t = 0$, we obtain an (electro-)
vacuum solution, which is presented in Appendix \ref{static} -- it is a
coordinate transform of the well-known Robinson solution \cite{Robi1959}. Thus
$C,_r = 0$, which means that the dust is moving on geodesics. A transformation
of $t$ can then be used to achieve $C = 0$. The other equations imply
\begin{equation}\label{3.2}
{R,_t}^2 = - 1 + \frac {2M} R - \frac {Q^2} {R^2} - \frac 1 3 \Lambda R^2,
\end{equation}
where $M$ is a constant of integration, and $Q$ is the electric charge within
the sphere of coordinate radius $r$. Since all other quantities in this equation
are independent of $r$, it follows that $Q$ must be constant. This implies that
the charge density is zero everywhere. Thus, the dust particles must be neutral,
they only move in an exterior electric field.

For the other metric function we obtain the following solution:
\begin{equation}\label{3.3}
{\rm e}^{A / 2} = R,_t \left[X(r) \int \frac {{\rm d} t} {R {R,_t}^2} +
Y(r)\right],
\end{equation}
where $X(r)$ and $Y(r)$ are arbitrary functions, and the expression for matter
density in energy units is
\begin{equation}\label{3.4}
\kappa \epsilon = \frac {2X} {R^2 {\rm e}^{A/2}}.
\end{equation}
Equations (\ref{3.1}) -- (\ref{3.4}) define the charged Ruban solution, first
semi-published in 1972 \cite{Ruba1972}, and then mentioned in a later paper
\cite{Ruba1983}. Note that (\ref{3.2}) is identical to (\ref{2.6}), so the
function $R(t)$ defined by (\ref{3.2}) will be the same as $R(\tau)$ defined by
(\ref{2.6}). Note also the following:

1. With $X = 0$, eq. (\ref{3.4}) shows that the Ruban solution becomes
electro-vacuum; in fact it then becomes the Reissner -- Nordstr\"{o}m solution
expressed in the $(\tau, \rho)$ coordinates defined by (\ref{2.5}) --
(\ref{2.6}).

2. When $Y = BX$, where $B$ is a constant ($B = 0$ being allowed), the
$r$-dependence in (\ref{3.4}) cancels out and by defining a new $r$ by $r' =
\int X(r) {\rm d} r$ we make also the metric independent of $r$. The spacetime
then becomes spatially homogeneous with the Kantowski -- Sachs symmetry; in fact
it is then the generalisation of the Kantowski -- Sachs solution to nonzero
charge and cosmological constant.

3. The geometry of a 3-space $t =$ const in (\ref{3.1}) is that of a
3-dimensional cylinder whose sections $r =$ const are spheres, all of the same
radius, and the coordinate $r$ measures the position along the generator. The
space is inhomogeneous along the $r$-direction, and the electric field has its
only component also in the $r$-direction. The radius $R$ of the cylinder evolves
with time according to Eq. (\ref{3.2}).

4. The matter density in this solution, given by (\ref{3.4}), depends on $r$ and
is everywhere positive if $X > 0$. Thus, the amount of rest mass contained
inside a sphere $r = r_0 =$ const does depend on the value of $r_0$, and is an
increasing function of $r$. Nevertheless, as seen from (\ref{3.2}), the active
gravitational mass $M$ that drives the evolution is constant. Ruban
\cite{Ruba1969} interpreted this property as follows: the gravitational mass
defect of any matter added exactly cancels its contribution to the active mass.

\section {Matching the Ruban and RN solutions}\label{matching}

\setcounter{equation}{0}

To match two spacetime regions to each other we have to prove that on the
hypersurface $H$ forming the border between them both 4-dimensional metrics
induce the same 3-dimensional metric and the same second fundamental form. The
formulae we use here are derived in Ref. \cite{PlKr2006}.

We will show that the RN metric in the coordinates of (\ref{2.6}) -- (\ref{2.7})
and the Ruban metric defined by (\ref{3.1}) -- (\ref{3.3}) with $C = 0$ can be
matched along any $H$ of constant $\rho = r$. On each such $H$ the 3-metrics
will be identical when $\tau$ and $t$ are identified and
\begin{equation}\label{4.1}
m = M, \qquad e = Q,
\end{equation}
since then the $R(\tau)$ of (\ref{2.7}) and the $R(t)$ of (\ref{3.1}) will obey
the same equation, (\ref{2.6}) and (\ref{3.2}) respectively.

The coincidence of the second fundamental forms requires that at $H$
\cite{PlKr2006}
\begin{equation}\label{4.2}
\left.\frac 1 {\cal N} \ \dr {h_{ij}} r\right|_{\rm Ruban} = \left.\frac 1 {\cal
N}\ \dr {h_{ij}} {\rho}\right|_{\rm RN},
\end{equation}
where $i, j = 0, 2, 3$ and ${\cal N}$ is the $r$-component of the unit normal
vector to $H$, thus $\left|g_{rr}\right| {\cal N}^2 = 1$ on each side of $H$.
But the relevant components of the Ruban metric do not depend on $r$, and the
relevant components of the RN metric do not depend on $\rho$, so (\ref{4.2}) is
fulfilled in a trivial way: it reduces to $0 = 0$. Note, however, that the
matching is possible only in that region of the RN manifold where $F < 0$ in
(\ref{2.1}) -- (\ref{2.2}). In the subcase $\Lambda = 0$, this is the region
between the two event horizons.

\begin{figure}
 \begin{center}
 \hspace*{20mm}
 \includegraphics[scale=0.6]{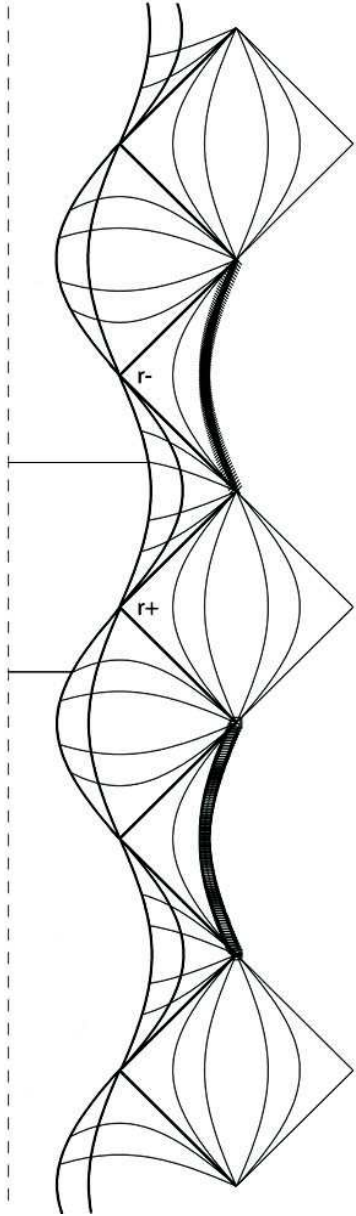}
 \caption{
 \label{rnrubanmatch}
 \footnotesize
The maximally extended Reissner--Nordstr\"{o}m spacetime with a hypothetical
charged matter source matched to it. The extent of the source in this diagram is
unknown, even if it is the charged Ruban solution -- because of the arbitrary
functions it contains. This is why the left edge is schematically marked with a
dashed line. The two long wavy curves show possible matchings at two different
values of the radial coordinate; this coordinate is the $r$ of (\ref{5.1}) and
the $\rho$ of (\ref{2.7}). Note: these are the curves on which $r$ and $\rho$
are constant. It is $R$ that pulsates between the $r_-$ and $r_+$ of (\ref{2.3})
and (\ref{2.4}), and it changes \textit{along} these curves. The lines of
constant $R$ are the horizontal hyperbola-like arches. If the matter source is
the charged Ruban solution, then shell crossings appear within each half-cycle
of oscillation. The positions of two of them are schematically marked with thin
horizontal straight lines.}
 \end{center}
 \end{figure}

Equation (\ref{2.9}) applies also in the Ruban region, with $m = M$ and $e = Q$.
It is independent of $r$, so each constant-$r$ shell evolves by the same law.
This means that all shells, including the outer surface of the Ruban region,
oscillate between $R = r_-$ and $R = r_+$, where $r_-$ and $r_+$ are given by
(\ref{2.3}) -- (\ref{2.4}). If the solution could be continued to infinite
values of $t$, the conformal diagram of the complete manifold would look as in
Fig. \ref{rnrubanmatch}. However, shell crossings make the completion of even
half a cycle of oscillations impossible, see next section.

\section {Shell crossings in the charged Ruban solution are inevitable}

\setcounter{equation}{0}

For the rest of this article we assume $\Lambda =0$. In this case the Ruban
metric is
\begin{equation}\label{5.1}
{\rm d} s^2 = {\rm d} t^2 - {\rm e}^{A(t, r)} {\rm d} r^2 - R^2(t)\left({\rm d}
\vartheta^2 + \sin^2 \vartheta {\rm d} \varphi^2\right),
\end{equation}
where the solution for $R(t)$ can be written as (cf. eq. (\ref{2.9})):
\begin{equation}\label{5.2}
R = M - \sqrt{M^2 - Q^2} \cos \eta, \qquad t - t_0 = \mu \left(M \eta -
\sqrt{M^2 - Q^2} \sin \eta\right),
\end{equation}
where $\mu = + 1$ in the expansion phase and $\mu = -1$ in the collapse phase.
With $\Lambda = 0$, the function ${\rm e}^{A/2}$ can be calculated explicitly.
For this purpose, we find from (\ref{3.2})
\begin{equation}\label{5.3}
R,_t = \mu \sqrt{- 1 + \frac {2M} R - \frac {Q^2} {R^2}}.
\end{equation}
Note that this solution exists only when $Q^2 < M^2$.\footnote{With $Q^2 = M^2$,
Eq. (\ref{5.3}) has the static solution $R = M = |Q|$. As remarked above
(\ref{3.2}), this leads to the Robinson solution \cite{Robi1959}; see Appendix
\ref{static}.} Now we use this in the integral in (\ref{3.3}) rewritten as
follows
\begin{equation}\label{5.4}
\int \frac {{\rm d} t} {R {R,_t}^2} \equiv \int \frac {{\rm d} R} {R {R,_t}^3}.
\end{equation}
Then we obtain
\begin{eqnarray}\label{5.5}
{\rm e}^{A/2} &=& X(r) \left[2 + Q^2 \frac {1 - M / R} {M^2 - Q^2} -\sqrt{- 1 +
\frac {2M} R - \frac {Q^2} {R^2}} \arcsin \left(\frac {R - M} {\sqrt{M^2 -
Q^2}}\right) \right] \nonumber \\
&+& \mu Y(r) \sqrt{- 1 + \frac {2M} R - \frac {Q^2} {R^2}}.
\end{eqnarray}

A shell crossing is where ${\rm e}^{A/2} = 0$, because at such a location two
matter shells whose $r$-coordinates differ by ${\rm d} r$ are at a zero
distance, as seen from (\ref{5.1}), and stick together. From (\ref{3.4}) one
sees that at such a location the mass density becomes infinite, so this is a
curvature singularity. The question we seek to answer is now: can the functions
$X(r)$ and $Y(r)$ in (\ref{5.5}) be chosen in such a way that ${\rm e}^{A/2}
\neq 0$ everywhere.

Unfortunately, the answer given by the formulae above is a definitive ``no'',
i.e. shell crossings are inevitable. To see this, let us first recall that $R$
changes between the values
\begin{equation}\label{5.6}
R_- = M - \sqrt{M^2 - Q^2}, \qquad {\rm and} \qquad R_+ = M + \sqrt{M^2 - Q^2}
\end{equation}
(this is a copy of (\ref{2.3}) -- (\ref{2.4}) rewritten in the notation for the
Ruban model). At both these values the expression under the square root in
(\ref{5.5}) is zero, and in the whole range $0 < R_- < R < R_+$ the function
${\rm e}^{A/2}$ given by (\ref{5.5}) is continuous. Now, at $R = R_-$ we have
\begin{equation}\label{5.7}
\left.{\rm e}^{A/2}\right|_{R = R_-} = - X(r) \frac {R_-} {\sqrt{M^2 - Q^2}},
\end{equation}
while at $R = R_+$ we have
\begin{equation}\label{5.8}
\left.{\rm e}^{A/2}\right|_{R = R_+} = X(r) \frac {R_+} {\sqrt{M^2 - Q^2}}.
\end{equation}
Thus, whatever the sign of $X(r)$, the signs of ${\rm e}^{A/2}$ at the ends of
the range $\left[R_-, R_+\right]$ are opposite, i.e. ${\rm e}^{A/2} = 0$ for
some value of $R$ within this range. This is a shell crossing. Note that it
appears every time when $R$ traverses this range, which means that the shell
crossings will not allow $R$ to go through even half a cycle of oscillation.

The geometrical nature of this shell crossing is different than in the LT and
Szekeres models \cite{Szek1975a,Szek1975b}, where they have been investigated
quite thoroughly \cite{HeLa1985,HeKr2002,HeKr2008}. In the LT and Szekeres
models, the spheres that collide are one within the other. At a shell crossing,
the smaller sphere catches up with the larger one during expansion (or vice
versa during collapse). In the Ruban model, the spheres are surfaces of constant
$r$ in the 3-dimensional cylinder $t =$ constant, and they move up and down
along the generators as the cylinder expands or collapses. It turns out they
will collide before the cylinder manages to proceed all the way from the minimal
radius to maximal, or from maximal to minimal. One more property of such a shell
crossing is noteworthy: its locus does not depend on $r$, which means that all
the spheres with different values of $r$ collide at the same moment.

Kantowski and Sachs \cite{KaSa1966} noted the possible occurrence of this
singularity in the subcase $Q = Y(r) = 0$ of the Ruban model, where they said
``The singularities for the closed models are of two kinds. (...) In one kind of
a singularity the cylinder squashes to a disk, in the other it contracts to a
line.'' What we called shell crossing here is the ``squashing to a disk''.

The shell crossings could possibly be prevented by pressure with a nonzero
gradient in the $r$-direction, but such exact solutions are not yet known. Then,
assuming that the pressure and its gradient would be zero at the surface of the
charged fluid ball, its surface would follow a timelike geodesic in the RN
spacetime, and the conformal diagram could really look like Fig.
\ref{rnrubanmatch}.

\section{Absence of shell crossings in the Ruban solution with zero
charge}\label{rubannocharge}

\setcounter{equation}{0}

When $Q = 0$, the charged Ruban solution goes over into the Datt -- Ruban
neutral dust solution \cite{Datt1938,Ruba1969,PlKr2006}. Then from (\ref{2.9}),
adapted to the notation for the Ruban solution, we obtain
\begin{equation}\label{6.1}
R = M (1 - \cos \eta), \qquad t - t_0 = \mu M (\eta - \sin \eta).
\end{equation}
This shows that now $R$ starts from 0 at $t = t_0$, then increases to $2M$ at $t
= t_0 + \mu M \pi$, and decreases to 0 again at $t = t_0 + 2 \mu M \pi$. At $R =
0$ the model has a Big Bang/Crunch type singularity. Equation (\ref{5.5})
simplifies to\footnote{Equation (\ref{6.2}) is equivalent to (19.101) in Ref.
\cite{PlKr2006} under the renaming $Y = \mu (\widetilde{Y} - \pi/2)$, in virtue
of the identity $\arcsin \left(R / M - 1\right) + \pi / 2 \equiv 2 \arcsin
\sqrt{R / (2M)}$.}
\begin{equation}\label{6.2}
{\rm e}^{A/2} = \sqrt{- 1 + \frac {2M} R} \left\{X(r) \left[\frac 2 {\sqrt{- 1 +
\frac {2M} R}} - \arcsin \left(\frac R M - 1\right)\right] + \mu Y(r)\right\}
\df \sqrt{- 1 + \frac {2M} R} {\cal F}(t,r).
\end{equation}
Now the spacetime is singular at $R = 0$, so the model exists only for the
finite time $T = 2 \pi M$ between the Big Bang and Big Crunch, but, unlike in
the charged case, the functions $X(r)$ and $Y(r)$ can be chosen so that within
this interval there are no shell crossings. We show below that, with an
appropriate choice, ${\cal F} > 0$ during the whole evolution.

Namely, with $R = 0$ at the start of the expansion phase, where $\mu = +1$, we
have
\begin{equation}\label{6.3}
\left.{\cal F}\right|_{R = 0} = X \frac {\pi} 2 + Y.
\end{equation}
Thus, $X > 0$ and $Y > 0$ will guarantee that $\left.{\cal F}\right|_{R = 0, \mu
= +1} > 0$. We also have
\begin{equation}\label{6.4}
\dr {\cal F} R = \frac X {R \left(\frac {2M} R - 1\right)^{3/2}},
\end{equation}
which will be positive for all $0 \leq R < 2M$ if $X > 0$, becoming $+ \infty$
as $R \to 2M$. Since also ${\cal F} \to + \infty$ as $R \to 2M$, the switch from
expansion to collapse, i.e. from $\mu = +1$ to $\mu = -1$, will keep ${\cal F}$
positive at the beginning of collapse. So we only have to guarantee that ${\cal
F} > 0$ at the end of the collapse phase, when $R \to 0$ again. This will be the
case when
\begin{equation}\label{6.5}
\left.{\cal F}\right|_{R = 0, \mu = -1} = X \frac {\pi} 2 - Y > 0.
\end{equation}
Consequently, both (\ref{6.3}) and (\ref{6.5}) will be positive during the whole
cycle when
\begin{equation}\label{6.6}
0 < Y < X \frac {\pi} 2
\end{equation}
at all values of $r$. The fact that the shell crossings become absent when $Q =
0$ shows that the limiting transition $Q \to 0$ is discontinuous; just as it was
in the RN $\to$ Schwarzschild transition.

The neutral Ruban solution can be matched to the Schwarzschild solution, as
shown by Ruban himself \cite{Ruba1969}. The matching hypersurface stays within
the Schwarzschild radius except at $R = 2M$, when it touches the horizon from
inside. However, this configuration has a finite time of existence, just as the
Kruskal \cite{Krus1960} -- Szekeres \cite{Szek1960} manifold between its two
singularities.

\section{Summary}\label{summ}

\setcounter{equation}{0}

We investigated the charged Ruban solution \cite{Ruba1972} as a possible matter
source for the maximally extended Reissner -- Nordstr\"{o}m solution with $e^2 <
m^2$. The matching of these two solutions is easily achieved. The hypersurface
that forms the boundary between the Ruban and RN regions stays all the time
between the two RN event horizons, touching the inner one at its minimal radius
and the outer one at maximal radius. However, shell crossings make the
completion of even half a cycle of such an oscillation impossible; they appear
between each $[r_-, r_+]$ and each $[r_+, r_-]$ pair.

Shell crossings can be avoided in the limit of zero charge, when the arbitrary
functions in the Ruban solution obey a simple inequality. Then the Ruban
solution can be matched to the Schwarzschild solution inside the event horizon,
but the model has a finite time of existence.

The problem of finding a matter source to the maximally extended RN solution
thus still remains open, with one more possibility being now elliminated.

\appendix

\section{The limiting case $Q^2 = M^2, \Lambda = 0$}\label{static}

\setcounter{equation}{0}

As stated in the footnote below (\ref{5.3}), this leads to the static solution
$R = M = |Q|$. This limit is not a subcase of (\ref{3.3}), and the Einstein --
Maxwell equations have to be solved anew. With $R,_t = 0$ they lead to the
electro-vacuum solution, in which
\begin{equation}\label{a.1}
{\rm e}^{A/2} = \alpha(r) \cos (t/M) + \beta(r) \sin (t/M),
\end{equation}
where $\alpha(r)$ and $\beta(r)$ are arbitrary functions, and the metric is
\begin{equation}\label{a.2}
{\rm d} s^2 = {\rm d} t^2 - {\rm e}^A {\rm d} r^2 - M^2\left({\rm d} \vartheta^2
+ \sin^2 \vartheta {\rm d} \varphi^2\right)
\end{equation}
(we recall that $M = |Q|$, where $Q$ is the electric charge). This should be
compared to the Nariai solution \cite{Nari1950} in the form found by Krasi\'nski
and Pleba\'nski \cite{KrPl1980}:
\begin{equation}\label{a.3}
{\rm d} s^2 = [a(t) \cos (\rho/\ell) + b(t) \sin (\rho/\ell)]^2 {\rm d} t^2 -
{\rm d} \rho^2 - \ell^2\left({\rm d} \vartheta^2 + \sin^2 \vartheta {\rm d}
\varphi^2\right),
\end{equation}
where $\ell$ is an arbitrary nonzero constant. This is a vacuum solution with
the cosmological constant $\Lambda = 1/\ell^2$. Equation (\ref{a.3}) is related
to (\ref{a.1}) -- (\ref{a.2}) by the interchange of $t$ and $r$, and has the
same geometrical structure -- it is a Cartesian product of two surfaces of
constant curvature.

In fact, (\ref{a.1}) -- (\ref{a.2}) is a coordinate transform of the Robinson
solution \cite{Robi1959}. Namely, by a rather complicated sequence of coordinate
transformations, strictly analogous to those used in Ref. \cite{KrPl1980} for
the Nariai solution, (\ref{a.1}) -- (\ref{a.2}) may be transformed to
\begin{equation}\label{a.4}
{\rm d} s^2 = \frac {\ell^2} {{r'}^2} \left[{\rm d} t^2 - {\rm d} {r'}^2 -
{r'}^2\left({\rm d} \vartheta^2 + \sin^2 \vartheta {\rm d}
\varphi^2\right)\right],
\end{equation}
which is identical with one of Robinson's formulae, with Robinson's $\lambda =
1/\ell$.

\bigskip

{\bf Acknowledgement:} The research of AK was supported by the Polish Ministry
of Education and Science grant no N N202 104 838. While the paper was being
prepared, GG was a summer intern at the N. Copernicus Astronomical Center in
Warsaw.

\end{document}